\documentclass[preprint,amsmath,amssymb,prd,superscriptaddress]{revtex4}
\usepackage[dvips]{graphicx}
\usepackage{color}
\usepackage{multirow}
\usepackage{array}
\usepackage{amsmath,bm}
\newcommand{\p}{\partial}
\newcommand{\pslash}{p\kern-1ex /}
\newcommand{\lslash}{l\kern-1ex /}
\newcommand{\kslash}{k\kern-1ex /}
\newcommand{\dslash}{\p\kern-1.2ex /}
\newcommand{\Dslash}{{\cal D}\kern-1.5ex /}
\newcommand{\Tr}{{\rm Tr}}
\newcommand{\tr}{{\rm tr}}

\newcommand{\bea}{\begin{eqnarray}}
\newcommand{\eea}{\end{eqnarray}}

\newcommand{\nn}{\nonumber\\}

\newcommand{\BAN}{\begin{eqnarray*}}
\newcommand{\EAN}{\end{eqnarray*}}
\begin{document}

\newcommand{\NTNU}{
  Physics Department, National Taiwan Normal University, Taipei, Taiwan~11677, R.O.C.
}

\newcommand{\ASIOP}{
  Institute of Physics, Academia Sinica, Taipei, Taiwan~11529, R.O.C. 
}

\newcommand{\NCTS}{
  Physics Division, National Center for Theoretical Sciences,
  National Tsing-Hua University, Hsinchu, Taiwan~30013, R.O.C.
}

\newcommand{\NTU}{
  Physics Department, National Taiwan University, Taipei, Taiwan~10617, R.O.C.
}

\newcommand{\CQSE}{
  Center for Quantum Science and Engineering,
  National Taiwan University, Taipei, Taiwan~10617, R.O.C. 
}

\newcommand{\RCAS}{
  Research Center for Applied Sciences, Academia Sinica,
  Taipei~11529, R.O.C.
}

\preprint{NTUTH-19-505B}

\title{Topological susceptibilty in lattice QCD with exact chiral symmetry -- 
       the index of overlap-Dirac operator versus the clover topological charge in Wilson flow   
} 

{

\author{Ting-Wai~Chiu}
\affiliation{\NTNU}
\affiliation{\ASIOP}
\affiliation{\NTU}

\author{Tung-Han~Hsieh}
\affiliation{\RCAS}

\collaboration{TWQCD Collaboration}
\noaffiliation


\begin{abstract}

Using an ensemble of 535 gauge configurations 
(on the $24^4 \times 48 $ lattice with $ a \simeq 0.06 $~fm and $ M_{\pi} \simeq 260 $~MeV)  
which are generated by hybrid Monte Carlo (HMC) simulation 
of $N_f=2$ lattice QCD with the optimal domain-wall quark,
we compute the index of the overlap-Dirac operator, and also measure the clover 
topological charge in the Wilson flow, $Q_{\text{clover}}(t) $, 
by integrating the flow equation from $ t = 0 $ to $ t = 128 $ with $\delta t = 0.01 $. 
We observe that $Q_{\text{clover}}(t) $ of each configuration converges to a value close 
to an integer, and its nearest integer $Q_c(t) = \text{round} [Q_{\text{clover}}(t)] $ becomes 
invariant for $ t \ge t_c $, with the $ \max\{t_c \} \sim 77 $ for all 535 configurations. 
For each configuration, we compare the asymptotically-invariant $ Q_c $ 
with the index of overlap-Dirac operator at $t=0$. It turns out that there are 167 configurations with 
$Q_c \ne \text{index}(D_{o}) $, amounting to $31.2\%$ of the total 535 configurations.  
However, the histograms of $ Q_c $ and $ \text{index}(D_o) $ are almost identical.  
Consequently, the topological susceptibility using the asymptotically-invariant $ Q_c $ 
agrees with that using the index of overlap-Dirac operator at $ t=0 $.
This implies that the topological susceptibility in lattice QCD with exact chiral symmetry 
can be obtained from the asymptotically-invariant $ Q_c $ in the Wilson flow.

\end{abstract}
\maketitle


The vacuum of Quantum Chromodynamics (QCD) has a non-trivial topological structure.
The gauge invariance and cluster property require that the ground state
must be the $\theta$ vacuum, a superposition of ground states in all topological sectors,  
\BAN
|\theta \rangle = \sum_{n} e^{i\theta n} |n\rangle,
\EAN 
where $ n $ is the winding number, and the summation goes over all integer values of $n$.
(For a pedagogical discussion of the $\theta$-vacuum, 
see, e.g., Refs. \cite{Weinberg:1996kr,Srednicki:2007qs}). 
The topological susceptibility $ \chi_t $  
is the most crucial quantity to measure the topological fluctuations of the QCD vacuum,  
which plays the important role in breaking the $ U_A(1) $ symmetry, and resolving  
the puzzle why the flavor-singlet $ \eta'$ is much 
heavier than other non-singlet (approximate) Goldstone bosons.
Moreover, the temperature dependence of $ \chi_t $ in QCD is the
crucial input to the phenomenology of axion cosmology.   
Formally, $ \chi_t $ is defined as
\bea
\label{eq:chit}
\chi_{t} = \int d^4 x  \left< \rho(x) \rho(0) \right>,
\eea
where $ \rho(x) $ is the topological charge density
expressed in term of the matrix-valued field tensor $ F_{\mu\nu} $, 
\BAN
\label{eq:rho}
\rho(x) = \frac{\epsilon_{\mu\nu\lambda\sigma}}{32 \pi^2} \tr[ F_{\mu\nu}(x) F_{\lambda\sigma}(x) ]. 
\EAN
From (\ref{eq:chit}), it gives
\bea
\label{eq:chit_Qt}
\chi_t = \frac{\left< Q_t^2 \right>}{\Omega}, \hspace{4mm}
Q_t \equiv \int d^4 x \rho(x),
\eea
where $ \Omega $ is the 4-dimensional volume of the system, and
$ Q_t $ is the topological charge (which is an integer for QCD).
Thus, $ \chi_t $ can be measured by counting the number of
gauge configurations in each topological sector.

However, in lattice gauge theory, the space of gauge field is connected
and the notion of a topological sector is not well-defined. 
Moreover, it is difficult to extract $ \rho(x) $
and $ Q_t $ unambiguously from the gauge link variables, 
due to their rather strong short-distance fluctuations. If one measures the 
clover topological charge $ Q_{\text{clover}} $ of any lattice QCD gauge configuration, 
it most likely turns out to be quite different from an integer. 
Moreover, its nearest integer $ Q_c = \text{round}[Q_{\text{clover}}] $
is also unreliable, except for the very smooth configurations.   
There are many proposals to smooth the gauge configuration.
However, it is unclear whether any of them can capture 
the ``genuine" topology of a gauge configuration.

Recently, the continuous-smearing \cite{Narayanan:2006rf}
or equivalently the Wilson flow \cite{Luscher:2010iy} 
has been widely used for smoothing the gauge configuration.  
Given a gauge configuration $ A_\mu $, the Wilson flow amounts to solving 
the discretized form of the following equation with respect to the 
fictituous flow time $ t $ (in unit of $a^2$), 
\bea
\label{eq:flow_eq}
\frac{d B_\mu}{dt} = D_\nu G_{\nu \mu},
\eea
with the initial condition $ B_\mu |_{t=0} = A_{\mu} $,
where $ G_{\nu\mu} = \partial_\nu B_\mu - \partial_\mu B_\nu + [B_\nu, B_\mu] $, and
$ D_\nu G_{\nu\mu} = \partial_\nu G_{\nu\mu} + [ B_\nu, G_{\nu\mu} ] $.
As shown in Ref. \cite{Luscher:2010iy},
the Wilson flow is a process of averaging gauge field over a spherical region of root-mean-square
radius $ R_{rms} = \sqrt{8 t} $. 

Now the first question is what flow time $ t $ should be used to 
measure the clover topological charge $ Q_{\text{clover}}(t) $ on the lattice.  
The second question is whether $ Q_{\text{clover}}(t) $ can capture the ``genuine" topological 
charge of the gauge configuration. 

In this paper, we address these two questions in lattice QCD with exact chiral symmetry 
\cite{Kaplan:1992bt,Neuberger:1997fp,Narayanan:1994gw},  
in view of that the overlap Dirac operator \cite{Neuberger:1997fp} 
in a topologically non-trivial gauge field
possesses exact zero modes with definite chirality satisfying
the Atiyah-Singer index theorem
\BAN
\text{index}(D_o) = n_+ - n_- = Q_t, 
\EAN
where $ n_\pm $ is the number of exact zero modes with $\pm$ chirality. 
Thus the index of overlap-Dirac operator can serve as the  
``genuine" topological charge for any gauge configuration on the lattice. 

Writing the overlap Dirac operator as
\BAN
\label{eq:overlap}
D_{o} = m_0 \left( 1 + \gamma_5 \frac{H_w}{\sqrt{H_w^2}} \right),
\EAN
where $ H_w = \gamma_5 D_w $ is the standard Hermitian Wilson operator plus a negative
parameter $ -m_0 $ ($ 0 < m_0 < 2 $), then its index is
\BAN
\label{eq:index_overlap}
 \mbox{index}(D_o)
= \Tr \left[ \gamma_5 \left( 1 - \frac{D_{o}}{2m_0} \right) \right]
= -\frac{1}{2} \Tr \left( \frac{H_w}{\sqrt{H_w^2}} \right)
= n_+ - n_-
= Q_t,
\EAN
where $ \Tr $ denotes trace over Dirac, color, and site indices.

We use an ensemble of 535 gauge configurations  
which are generated by hybrid Monte Carlo (HMC) simulation 
of $N_f=2$ lattice QCD with optimal domain-wall quark \cite{Chiu:2002ir}, 
and Wilson plaquette gauge action,  
on the $24^4 \times 48 $ lattice with $ a \simeq 0.06 $~fm and $ M_{\pi} \simeq 260 $~MeV. 
The parameters for the HMC simulation are $ \beta = 6/g_0^2 = 6.10 $, 
$m_{u/d} = 0.005$, $N_s=16$, $\lambda_{min}/\lambda_{max} = 0.05/6.20$, and $m_0=1.3$. 
This ensemble is exactly the ensemble $A$ as listed in the Table I of Ref. \cite{Chen:2014hva}, 
for the first study of pseudoscalar decay constants $ f_D $ and $ f_{D_s} $ in $N_f=2$ lattice QCD 
with domain-wall fermion. More details of this gauge ensemble are given in Ref. \cite{Chen:2014hva}. 

For each configuration, we compute $ n_\pm $ zero modes 
and $(200-n_\pm)$ conjugate pairs of the lowest-lying eigenmodes of the overlap-Dirac operator.
We outline our procedures as follows.
First, we project 400 low-lying eigenmodes of $ H_w^2 $ using adaptive
thick-restart Lanczos algorithm ($a$-TRLan) \cite{a-TRLan},
where each eigenmode has a residual less than $ 10^{-12} $.
Then we approximate the sign function of the overlap operator
by the Zolotarev optimal rational approximation with 64 poles,
where the coefficients are fixed with $ \lambda_{max}^2 = (6.2)^2 $,
and $ \lambda_{min}^2 $ equal to the maximum of
the 400 projected eigenvalues of $ H_w^2 $.
Then the sign function error is less than $ 10^{-14} $.
Using the 400 low-modes of $ H_w^2 $ and the Zolotarev approximation
with 64 poles, we use the $a$-TRLan algorithm again to
project the $ n_\pm $ zero modes 
and $(200-n_\pm)$ conjugate pairs of the lowest-lying eigenmodes of the overlap-Dirac operator, 
where each eigenmode has a residual less than $ 10^{-12} $.
More details of our procedures are given in Refs. \cite{Chiu:2011dz, Chiu:2014hga}.
For each configuration, we use the index of the overlap Dirac operator as the
topological charge of this configuration ($Q_t = n_+ - n_- $),   
and obtain the topological susceptibility (\ref{eq:chit_Qt})
\bea
\label{eq:chit_overlap}
\chi_t a^4 = 7.03(91) \times 10^{-7}.
\eea 
The histogram of the probability distribution of $ \text{index}(D_o) $ 
is plotted in Fig \ref{fig:hist} (a).

\begin{figure*}[ht!]
\begin{center}
\begin{tabular}{@{}c@{}c@{}}
\includegraphics*[width=8cm,clip=true]{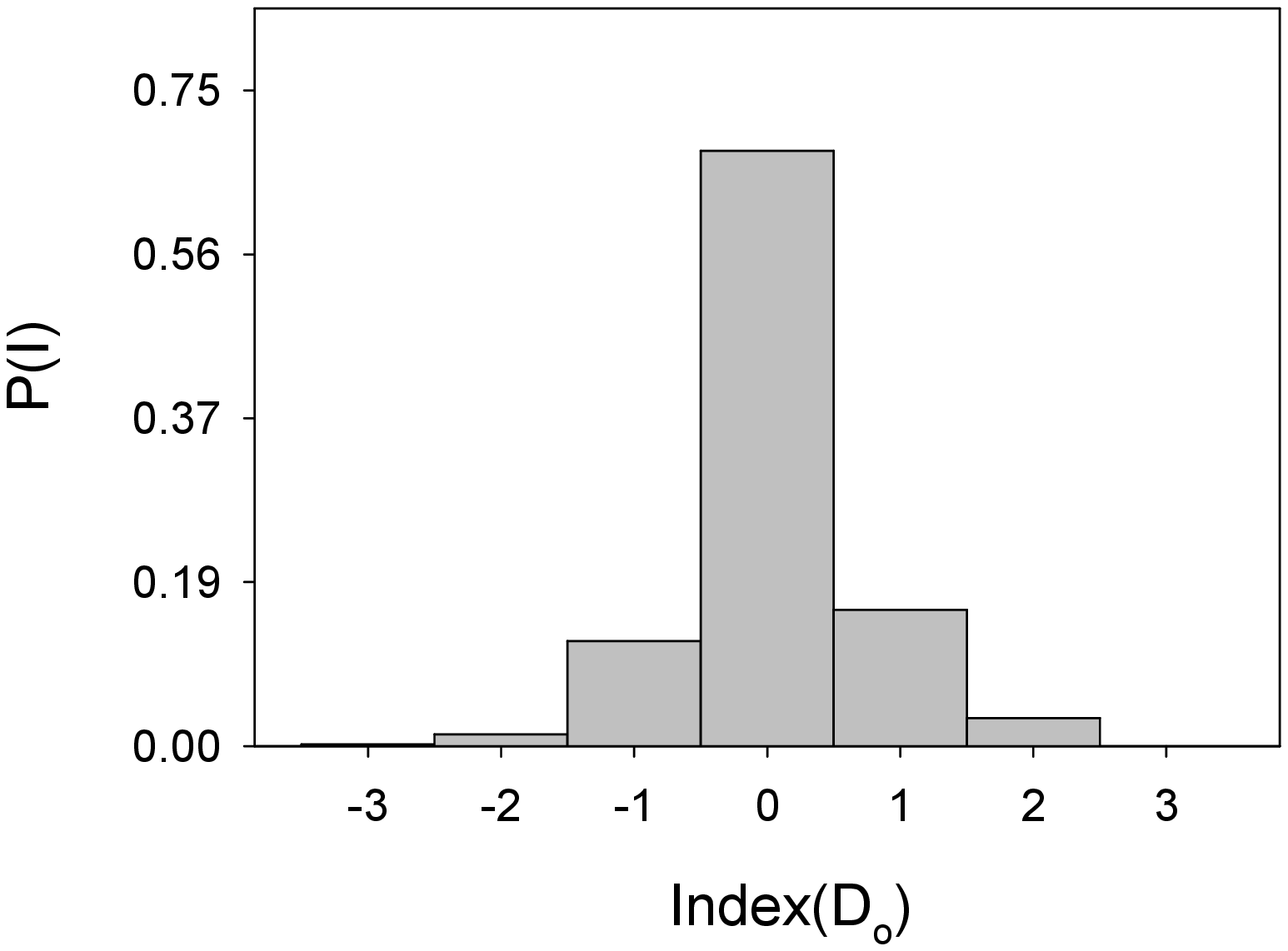}
&
\includegraphics*[width=8cm,clip=true]{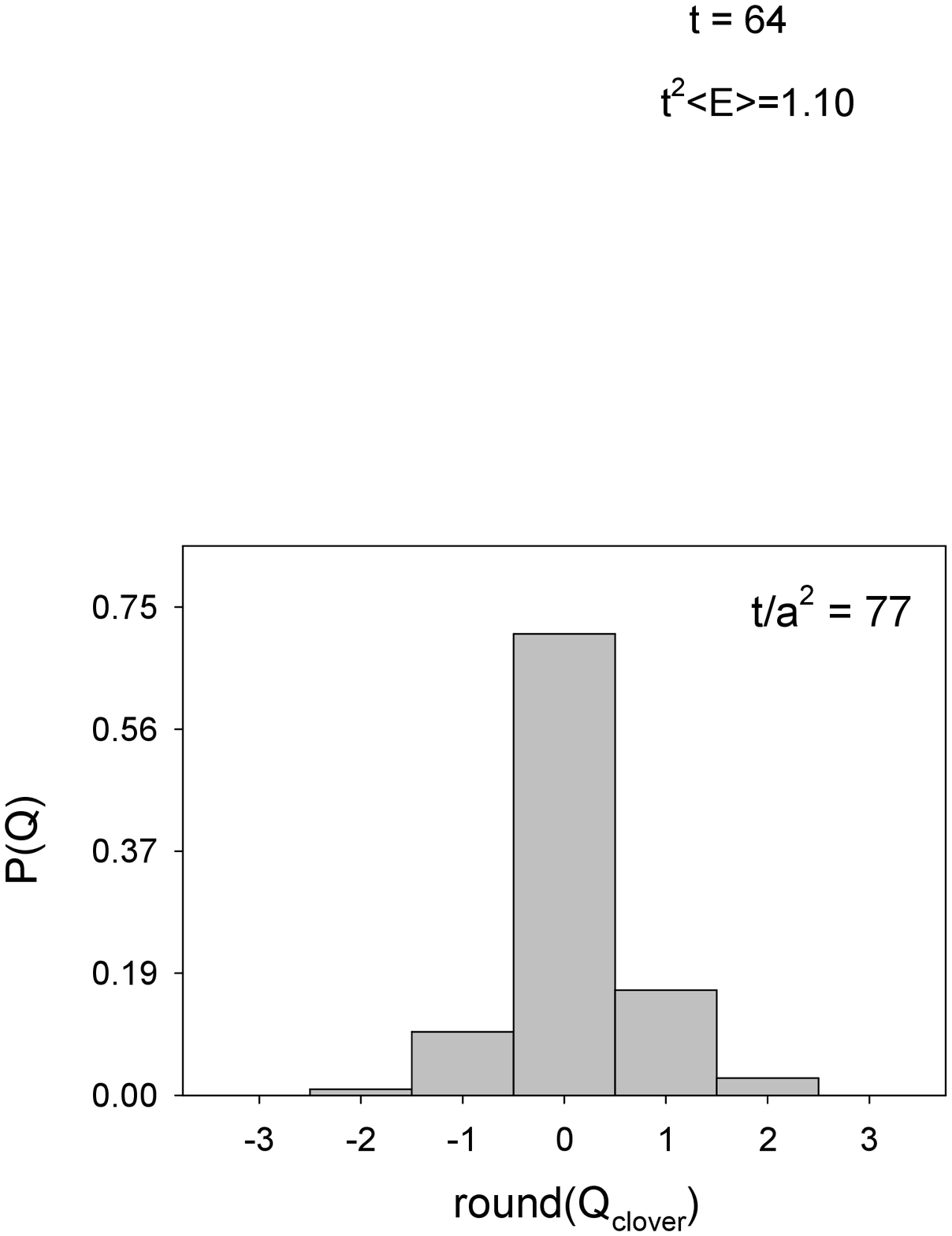}
\\ (a) & (b)
\end{tabular}
\caption{
Histogram of the probability distribution of topological charge 
of an ensemble of 535 configurations in $ N_f=2 $ lattice QCD with the optimal DWF.
(a) The index of the overlap-Dirac operator at $t=0$.
(b) The asymptotically-invariant clover topological charge $Q_{tc}$.  
}
\label{fig:hist}
\end{center}
\end{figure*}

Next we perform the Wilson flow by numerically integrating the discretized form 
of the flow equation (\ref{eq:flow_eq}) from $ t = 0 $ to $ t = 128 $ with $\delta t = 0.01 $, 
and also measure $Q_{\text{clover}}(t) $ along the Wilson flow, in which the  
the matrix-valued field tensor $ F_{\mu\nu}(x) $ entering (\ref{eq:rho}) is obtained from
the four plaquettes (clover) surrounding $ x $ on the $ (\hat\mu, \hat\nu) $ plane, i.e.,
\BAN
F_{\mu\nu}(x)
& \simeq & 
\frac{1}{8i}
   [  P_{\mu\nu}(x) + P_{\mu\nu}(x-\hat\mu) + P_{\mu\nu}(x-\hat\nu)
      + P_{\mu\nu}(x-\hat\mu-\hat\nu)  \nn
& &
\hspace{2mm}
      - P^{\dagger}_{\mu\nu}(x) - P^{\dagger}_{\mu\nu}(x-\hat\mu)
      - P^{\dagger}_{\mu\nu}(x-\hat\nu)
      - P^{\dagger}_{\mu\nu}(x-\hat\mu-\hat\nu) ], 
\EAN 
where 
$ P_{\mu\nu}(x) = V_\mu(x) V_\nu(x+\hat\mu) V^{\dagger}_\mu(x+\hat\nu) V^{\dagger}_\nu(x) $,
and $ V_\mu(x) $ denotes the link variable at the flow time $t$.

\begin{figure*}[ht!]
\begin{center}
\includegraphics*[width=10cm,clip=true]{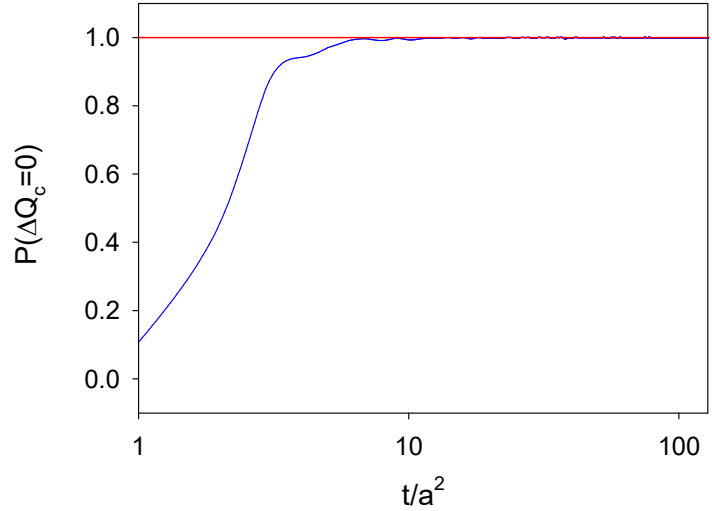}
\caption{
The fraction of the total 535 configurations with $ \Delta Q_c = 0 $ within the flow time interval 
$[t-\Delta t, t ]$ for $ \Delta t = 1 $, versus the Wilson flow time $t$. 
Note that the $t$-axis is in the common log scale.
}
\label{fig:NdQ=0}
\end{center}
\end{figure*}
 
We observe that $Q_{\text{clover}}(t) $ of each configuration converges to a value close 
to an integer (see the examples in Fig. \ref{fig:evolution}), 
and its nearest integer $Q_c(t) = \text{round} [Q_{\text{clover}}(t)] $ becomes 
invariant for $ t \ge t_c $, with the $ \max\{t_c\} \sim 77 $ for all 535 configurations. 
In Fig. \ref{fig:NdQ=0}, the fraction of the total 535 configurations with $ \Delta Q_c = 0 $ 
within the Wilson flow time interval $[t-1, t]$ is plotted versus the Wilson flow time $t$. 
Evidently, most configurations have attained the status with $ \Delta Q_c = 0 $ for  
$ t \gtrsim 10 $, however, there are still a tiny fraction ($ < 0.005$) of the total 535 configurations 
with $ \Delta Q_c \ne 0 $. Only after $ t \ge 77 $, 
the $ Q_c $ of each configuration becomes invariant.          
In other words, for $ t \ge \max\{t_c\} \sim 77 $, 
all 535 configurations become sufficiently smooth to decompose into topological sectors, 
similar to the gauge fields in the continuum theory.
Thus it is natural to use the asymptotically-invariant $ Q_c $ (denoted by $Q_{tc}$) 
of each configuration as its topological charge, 
and to compute the topological susceptibility $\chi_t $ (\ref{eq:chit_Qt}) with $Q_{tc}$.
This answers the first question what flow time $ t $ should be used to 
measure the clover topological charge $ Q_{\text{clover}}(t) $ on the lattice.

Recall that the condition for a lattice gauge configuration to fall into a topological sector 
has been discussed in Refs. \cite{Luscher:1981zq,Phillips:1986qd,Luscher:2010iy}.
For lattice QCD gauge configuration on the hypercubical lattice,   
it can be written as  
\bea
\label{eq:admissibility}
\min_{\forall x,\mu,\nu} \left\{\frac{1}{3}{\text{Re}} \ \tr \ P_{\mu\nu}(x) \right\} 
>\frac{44}{45} \simeq 0.978, 
\eea
where $ P_{\mu\nu}(x) $ is the ordered product of link variables around a plaquette. 
That is, if all plaquette values of a lattice QCD gauge configuration are kept to be greater than 0.978, 
then its topological charge would not be changed by continuous deformation of the gauge fields. 
Presumably, any smoothing algorithm can bring a configuration to satisfy (\ref{eq:admissibility}). 
The question is whether the resulting gauge configuration falls into the proper topological sector or not. 
For a gauge ensemble, the less restrictive but relevant question 
is whether the resulting ensemble of gauge configurations 
can capture the topological fluctuations of the QCD vacuum, which are  
measured by the topological susceptibiliy and the higher moments ($c_4, \cdots$) 
of the topological charge distribution.    

\begin{figure*}[ht!]
\begin{center}
\includegraphics*[width=10cm,clip=true]{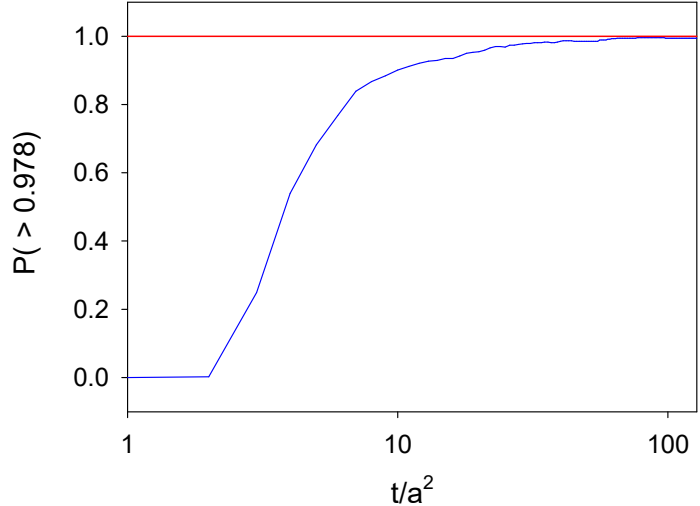}
\caption{
The fraction of the total 535 configurations satisying the condition (\ref{eq:admissibility})
versus the Wilson flow time $t$. 
Note that the $t$-axis is in the common log scale.
}
\label{fig:P0978}
\end{center}
\end{figure*}
   
Now it is interesting to check whether all 535 configurations in this ensemble 
satisfy the condition (\ref{eq:admissibility}) for $ t \ge t_c $. 
In Fig \ref{fig:P0978}, 
the fraction of the total 535 configurations satisfying the condition (\ref{eq:admissibility})
is plotted versus the Wilson flow time $t$. For $ t \ge \max\{t_c\} \sim 77 $, 
there are 3 configurations not satisfying (\ref{eq:admissibility}), 
amounting to $\sim 0.6\% $ of the total 535 configurations.
This implies that (\ref{eq:admissibility}) is not the necessary condition for  
a lattice QCD gauge configuration to fall into a topological sector, 
but a sufficient condition. This can also be seen by comparing Fig. \ref{fig:P0978} 
with Fig. \ref{fig:NdQ=0}. In Fig. \ref{fig:NdQ=0}, at $ t = 10 $, 
more than $99\%$ of the configurations have reached their asymptotically-invariant $ Q_{tc} $, 
and have fallen into topological sectors. On the other hand, in Fig. \ref{fig:P0978}, at $ t = 10 $, 
only about $90\%$ of the configurations satisfy the condition (\ref{eq:admissibility}).

Next we turn to the second question whether $ Q_{tc} $ can capture the ``genuine" 
topological charge of any lattice QCD gauge configuration.
Comparing $Q_{tc}$ with the $\text{index}(D_o) $ at $t=0$,  
we find that there are 167 configurations with 
$Q_c \ne \text{index}(D_{o}) $, amounting to $31.2\%$ of the total 535 configurations.  
In Fig. \ref{fig:evolution}, we present examples of two different cases: 
(a) $ Q_{tc} = \text{index}(D_o) $, and (b) $ Q_{tc} \ne \text{index}(D_o)$.
Now the questions are what causes the discrepancy between $ Q_{tc}$ and the $ \text{index}(D_o) $ at $t=0$, 
for $31.2\%$ configurations in this ensemble, and whether the discrepancy also manifests 
in the topological susceptibility.

\begin{figure*}[ht!]
\begin{center}
\begin{tabular}{@{}c@{}c@{}}
\includegraphics*[width=8cm,clip=true]{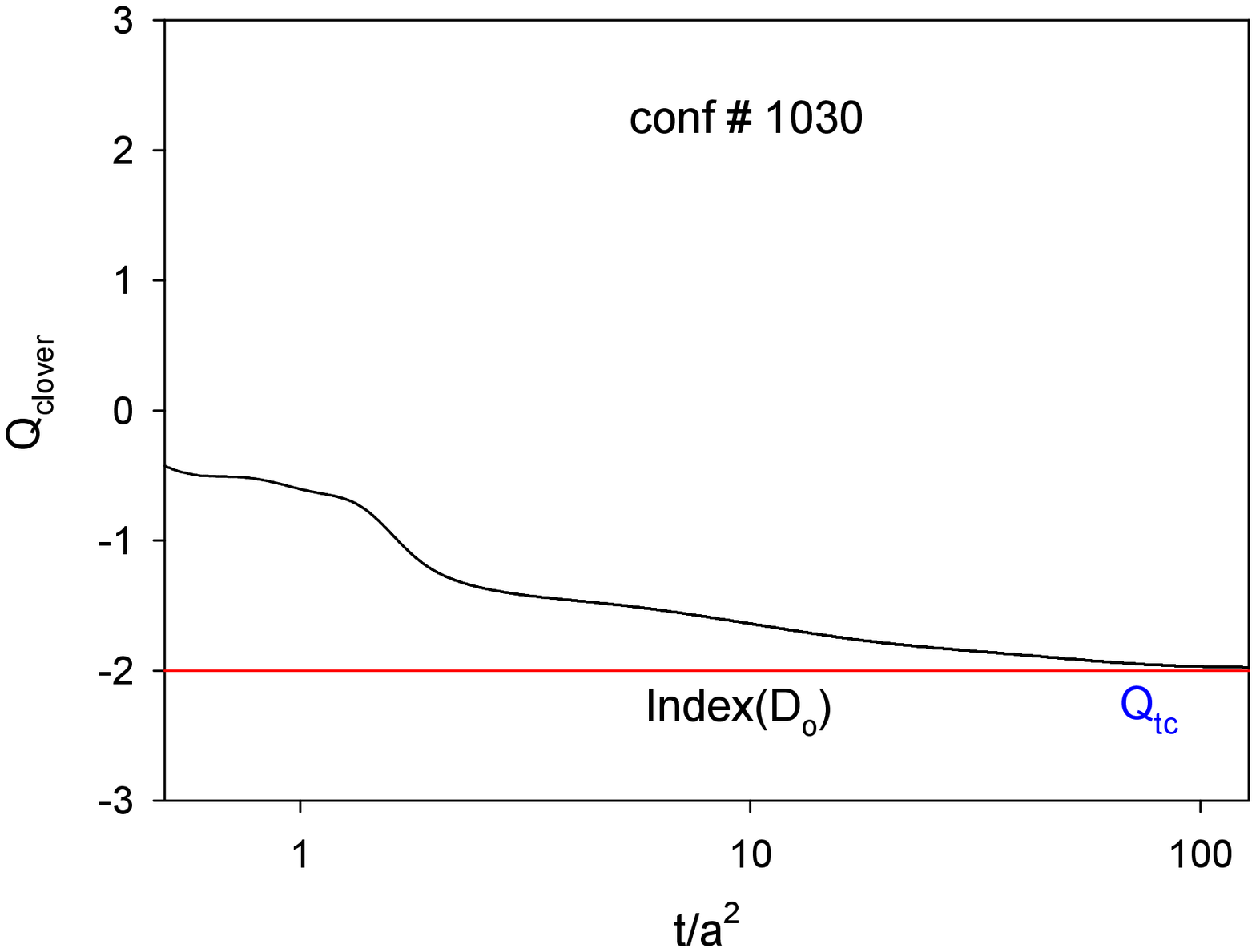}
&
\includegraphics*[width=8cm,clip=true]{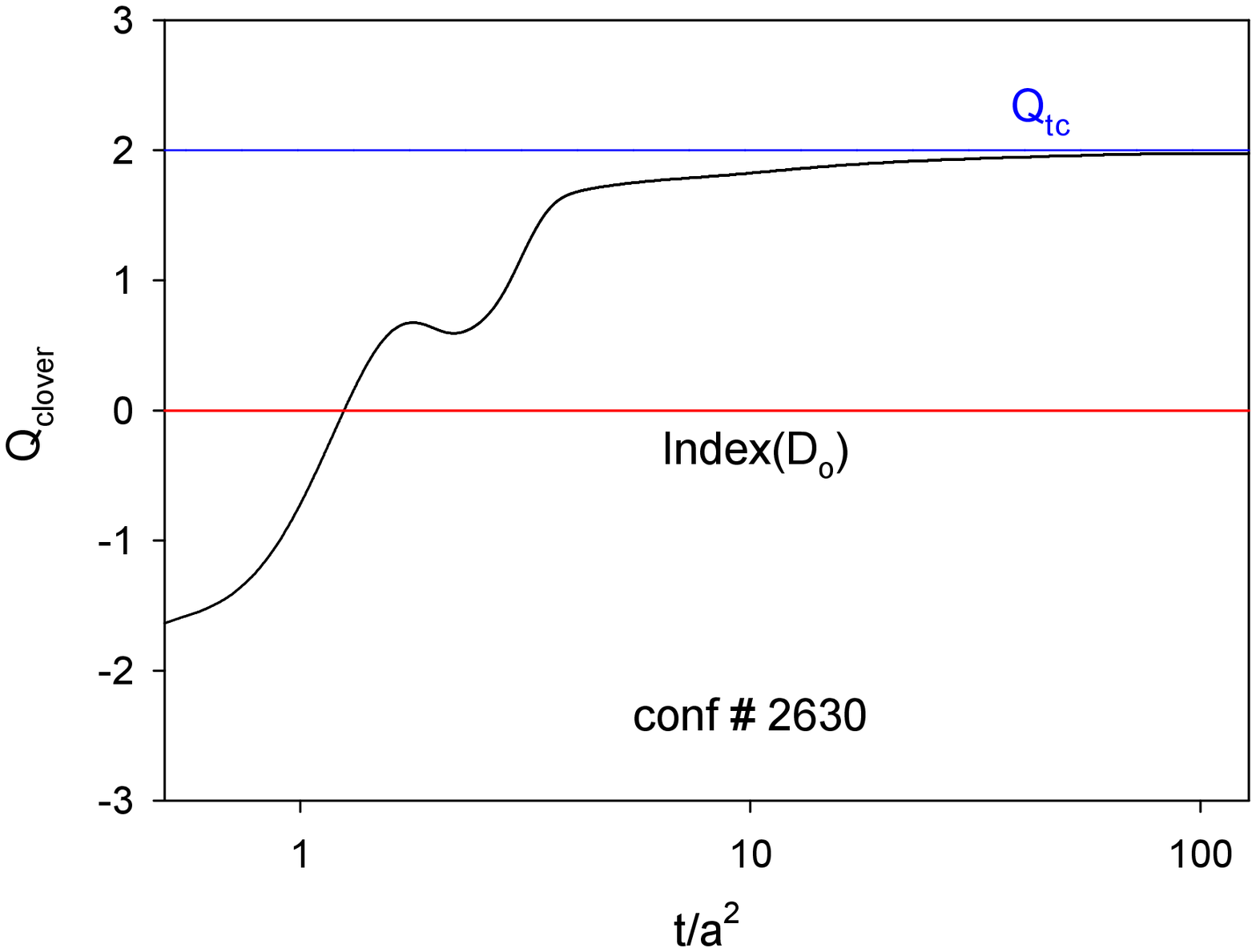}
\\ (a) & (b)
\end{tabular}
\caption{
Two examples of the evolution of the clover topological charge versus the flow time $t$.
(a) $ Q_{tc} = \text{index}(D_o) $.     
(b) $ Q_{tc} \ne \text{index}(D_o)$.
The horizontal line is the $\text{index}(D_o)$ of the gauge configuration at $ t = 0 $. 
}
\label{fig:evolution}
\end{center}
\end{figure*}

\begin{figure*}[ht!]
\begin{center}
\includegraphics*[width=10cm,clip=true]{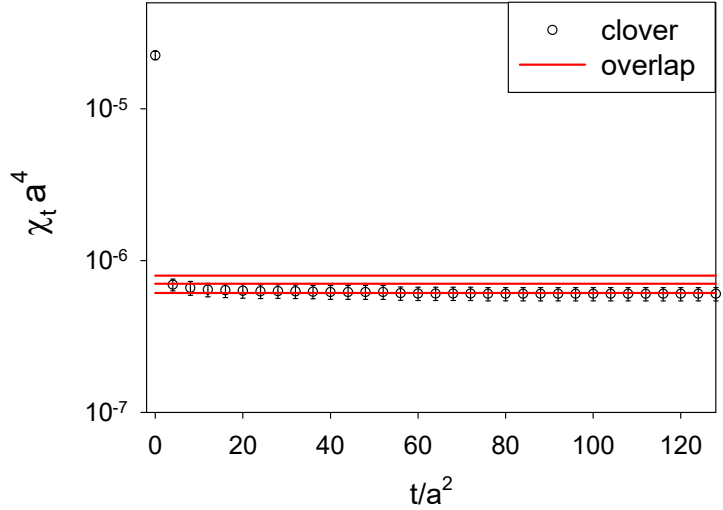}
\caption{
The topological susceptibility $ \chi_t $ of the 
rounded clover topological charge $Q_c(t)$ 
versus the flow time $ t $. The central horizontal line is the 
the topological susceptibility of the index of the overlap-Dirac operator at $ t = 0 $, 
with the error bar as the upper and the lower horizontal lines.
%
}
\label{fig:chit_t}
\end{center}
\end{figure*}

Since this gauge ensemble has decomposed into topological sectors after flowing    
for $ t \ge \max\{t_c\} \sim 77 $, the topological susceptibility $\chi_t $ (\ref{eq:chit_Qt})  
can be computed with $ Q_{tc} $, giving 
\bea
\label{eq:chit_wflow}
\chi_t a^4 = 6.03(61) \times 10^{-7}, 
\eea 
which agress with that (\ref{eq:chit_overlap}) using the index of overlap-Dirac operator at $t=0$.  
The histogram of the probability distribution of $ Q_{tc} $ is plotted in Fig \ref{fig:hist} (b), 
which is almost identical to that of the $\text{index}(D_o)$ at $ t = 0 $ in Fig. \ref{fig:hist} (a). 
It is interesting to see that  
the topological susceptibility using the asymptotically-invariant $ Q_c $ 
is in good agreement with that using the index of overlap-Dirac operator at $ t=0 $, 
even though $ Q_{tc} \ne \text{index}(D_o) $ for $31.2\%$ of the total configurations. 
This implies that the topological susceptibility in lattice QCD with exact chiral symmetry 
can be computed with the asymptotically-invariant $ Q_c $ in the Wilson flow. 

In Fig. \ref{fig:chit_t}, the topological susceptibility $\chi_t $ obtained with  
$ Q_c(t) = \text{round}[Q_{\text{clover}}(t)] $ is plotted versus the flow time $ t $.
It is interesting to see that $\chi_t$ attains a plateau starting from $ t \sim 10 $, 
long before all 535 configurations fall into topological sectors at $ t \sim 77 $. 
This is consistent with the scenario in Fig. \ref{fig:NdQ=0} that more than $99\%$ 
of the configurations in the ensemble have already reached 
the asymptotically-invariant $ Q_{tc} $ at $ t \sim 10 $.  
Now it seems tempting to extract $ \chi_t $ from its plateau in some interval
of the Wilson flow, e.g., $ 10 \le t \le 20 $,  
without worrying whether the flow has reached the $ \max\{t_c\} \sim 77 $ or not. 
Strictly speaking, this is not theoretically justified 
since there are still some configurations in the gauge ensemble have not fallen 
into the topological sectors yet.
Thus the eligible procedure is to perform the Wilson flow up to 
$ t = \max\{t_c\} $ for all configurations of the gauge ensemble 
such that all configurations are decomposed into topological sectors, 
then it is justified to compute $ \chi_t $ with $ \{ Q_{tc} \} $ of all configurations. 

Finally we return to the question what causes the discrepancy between 
$ Q_{tc}$ and the $ \text{index}(D_o) $ at $t=0$. 
In general, for any lattice gauge configuration (at $t=0$), 
a priori, one cannot prove whether the $ \text{index}(D_o) $ of this configuration is 
equal to the clover topological charge $ Q_{tc} $ 
(which is obtained from the sufficiently smooth configuration at $ t = t_c $) or not. 
However, one can ask whether the $ \text{index}(D_o) $ 
is equal to $ Q_{tc} $, for the same gauge configuration at $ t = t_c $.  
To answer this question, we project $ n_\pm $ zero modes 
and $(200-n_\pm)$ conjugate pairs of the lowest-lying eigenmodes of the overlap-Dirac operator 
with the gauge configuration at $t=77$, for all 535 configurations. 
We verify that $Q_{tc}$ is exactly equal to the $ \text{index}(D_o) $ at $t=77$, 
for each configuration of this ensemble. 
In other words, even if $ \text{index}(D_o) \ne Q_c = \text{round} [Q_{\text{clover}}] $  
for a rough gauge configuration at $t=0$, it evolves along the Wilson flow   
and eventually reaches the equality, $ \text{index}(D_o) = Q_c $,   
when the gauge configuration becomes sufficiently smooth for $ t \ge t_c $. 
Since the $\chi_t$ computed with the $\text{index}(D_o) $ at $ t=0 $ is in good agreement with
the $\chi_t$ computed with $Q_{tc}$ (as shown in Fig. \ref{fig:chit_t}),
and the $\text{index}(D_o)$ at $t=t_c$ is exactly equal to $Q_{tc}$, 
this implies that the $\chi_t$ computed with the $\text{index}(D_o) $
is almost invariant with respect to the Wilson flow time $t$, 
regardless of whether the gauge ensemble is sufficiently smooth or not. 
This is an appealing feature of the overlap-Dirac operator.

For lattice QCD with non-chiral fermions, it is unknown whether the asymptotically-invariant $Q_{tc}$ 
could exist in the Wilson flow or not, due to some uncontrollable lattice artifacts. 
Even if $ Q_{tc} $ exists for each configuration in a gauge ensemble, 
it is still uncertain whether the $ \chi_t $ 
and the higher moments ($c_4, \cdots$) computed with $ \{ Q_{tc} \} $
do capture the ``genuine" topological fluctuations of the QCD vacuum.   
Even for lattice QCD with exact chiral symmetry, the lattice artifacts may  
have sizable effects for lattice spacing $ a > 0.1 $~fm, which in turn would give a distorted picture 
different from what we have seen in this study using a gauge ensemble with $ a \sim 0.06 $~fm.

To summarize, using an ensemble of gauge configurations of $N_f=2$ lattice QCD 
with the optimal domain-wall quark (for which the effective 4-dimensional lattice Dirac operator 
is exactly equal to the Zolotarev optimal rational approximation of the overlap-Dirac operator),   
we compute the index of the overlap-Dirac operator, and also measure the clover 
topological charge in the Wilson flow, by integrating the flow equation from 
$ t = 0 $ to $ t = 128 $ with $\delta t = 0.01 $. 
We observe that the clover topological charge $Q_{\text{clover}}(t) $ of each configuration converges 
to a value close to an integer, and its nearest integer $Q_c(t) = \text{round} [Q_{\text{clover}}(t)] $ 
becomes invariant for $ t \ge t_c $, with the $ \max\{t_c \} \sim 77 $ 
for all configurations in the ensemble. 
This asserts that the gauge ensemble is decomposed into topological sectors for $ t \ge \max\{t_c\} $, 
similar to the gauge fields in the continuum theory, and also provides the guideline for computing 
the $\chi_t$ with the $Q_{\text{clover}}$ in the Wilson flow. 
Moreover, we find that the $ \chi_t $ computed with the asymptotically-invariant $ Q_c $ 
agrees with that using the $\text{index}(D_o)$ at $t=0$.    
This implies that the topological fluctuations of the QCD vacuum (i.e., $\chi_t, c_4, \cdots$)  
in lattice QCD with exact chiral symmetry can be obtained with $ Q_{tc} $ in the Wilson flow.

\begin{acknowledgments}

This work is supported by the Ministry of Science and Technology
(Grant Nos.~108-2119-M-003-005, 107-2119-M-003-008, 105-2112-M-002-016, 102-2112-M-002-019-MY3),
and the National Center for Theoretical Sciences (Physics Division).
We gratefully acknowledge the computer resources provided by
Academia Sinica Grid Computing Center (ASGC), and National Center for High Performance Computing (NCHC).  

\end{acknowledgments}

\end{document}